\documentclass[useAMS,usenatbib]{mn2e}
\usepackage{amsmath}
\usepackage{graphicx}

\newcommand{\beq}{\begin{equation}}
\newcommand{\eeq}{\end{equation}}

\begin{document}

\title[Probing the Spacetime Around Supermassive Black Holes with Ejected Plasma Blobs]{Probing the Spacetime Around Supermassive Black Holes with Ejected Plasma Blobs}

\author[P. Christian and A. Loeb]
{Pierre Christian$^1$ and 
Abraham Loeb$^1$ 
\\
$^1$Harvard-Smithsonian Center for Astrophysics, 60 Garden Street, Cambridge, MA\\}

\maketitle
\begin{abstract}
Millimeter-wavelength VLBI observations of the supermassive black holes in Sgr A* and M87 by the Event Horizon Telescope could potentially trace the dynamics of ejected plasma blobs in real time. We demonstrate that the trajectory and tidal stretching of these blobs can be used to test general relativity and set new constraints on the mass and spin of these black holes.
\end{abstract}

\section{Introduction}

The planned Event Horizon Telescope (EHT)\footnote{http://www.eventhorizontelescope.org/} will possess angular resolution comparable to the Schwarzschild radius of the supermassive black holes (SMBHs), Sgr A* and the one at the center of M87, and temporal resolution on minutes timescales \citep{Johnson}. This is expected to open a new avenue for studying a multitude of transient phenomenae under extreme gravity. 

Sgr A* is known to exhibit variability with tens of minutes timescale corresponding to accretion disk activity at the innermost stable circular orbit (ISCO) \citep{Genzel, Johnson}. Here we study a hypothetical class of short timescale events corresponding to plasma blobs ejected near the ISCO radius. Although such blobs were never observed from a supermassive black hole, they may exist based on the analogy with microquasars, which are known to propel blobs at relativistic speeds \citep{mirabel,mirabel2,mirabel3}. 

The second target of the EHT is the supermassive black hole at the center of the elliptical galaxy M87. In contrast to Sgr A*, M87 possesses a jet, and it is likely that blobs are ejected along the jet's symmetry axis.

In this \emph{Letter}, we demonstrate that if ejected plasma blobs were detected, one could use their dynamics to probe the spacetime around the black holes. Furthermore, if the mass and spin of a given black hole are known, one can use observations of the blob's dynamics to test general relativity or infer the presence of non-gravitational sources such as gas pressure or magnetic stress. These constraints would be complimentary to constraints from pulsar timing \citep{c4,c10,c11,c5,c6} or observations of the black hole shadow \citep{a1,a2,a3}. 

There are two elements of dynamical information: the trajectory of the blob's center of mass, and its lateral expansion. Both can be used to independently constraint the black hole's spacetime.
We discuss the former in \S 2 and \S 3, and the later in \S 4. Throughout the discussion, we will assume general relativity. Deviations from our results would indicate the presence of non-gravitational forces or corrections to the  theory of gravity. We use units where $G=c=1$, and the conversion from these units to physical units is given in Table \ref{Mval}.

\begin{table*}\label{Mval}
\begin{center} 
\begin{tabular}{ || c|c |c|c|c |c  || } 
 \hline
  & Black hole mass & Distance & Time & Space & Angle \\ 
 \hline
Sgr A* &$(4.31 \pm 0.36) \times 10^6 M_\odot$  & $7.94 \pm 0.42$ kpc & $21$ s & $0.043$ AU & $5.3$ $\mu$as \\ 
 M87 & $(6.6 \pm 0.4) \times 10^9 M_\odot$& $16.7 \pm 0.9$ Mpc & $9$ hr & $65$ AU  & $3.9$ $\mu$as  \\ 
 \hline
\end{tabular}
\caption{The conversion of black hole mass, $M$, to units of time, space, and angular size on the sky for Sgr A* and M87 \citep{massbh,ghez,M87,Dsgr,DM87}, for $G=c=1$.}
\end{center} 
\end{table*}

\section{Center of Mass Motion}

\begin{figure}
\centering
\includegraphics[width=0.4\textwidth]{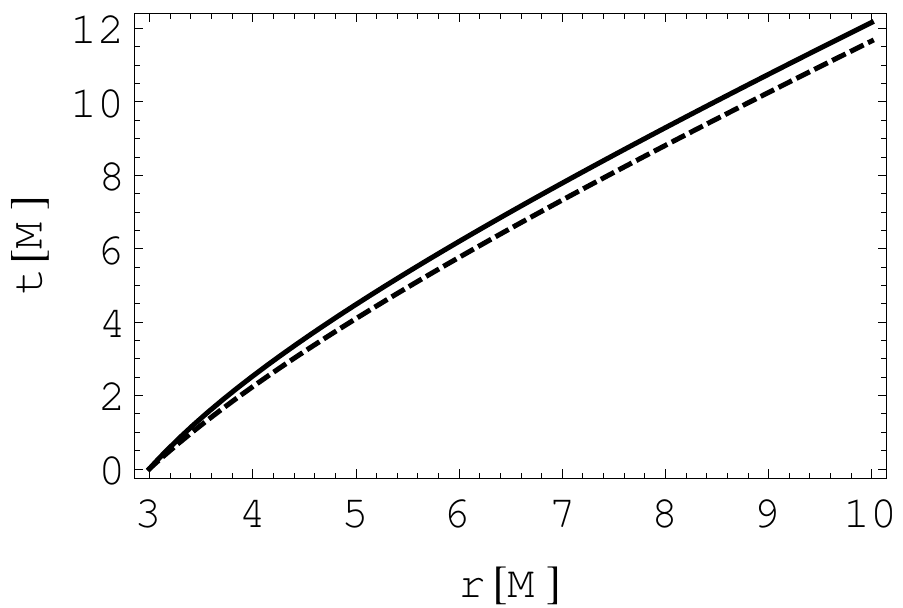}
\caption{The radial motion of blobs with $e=2$ in the equitorial plane of a black hole with $a=0$ (solid line) and $a=0.999$ (dotted line). }
\label{spin}
\end{figure}

First we consider the motion of the blob's center of mass (COM). If the blob is ejected above the escape speed from the ISCO radius, $R_{ISCO}$, its azimuthal velocity will be negligibly small at $r \gg R_{ISCO}$, so we focus our discussion on the radial equation of motion. For a Schwarzschild black hole \citep{Chandra},
\beq
\left(\frac{dr }{d \tau}\right)^2 = \frac{2 M }{r} - (1 - e^2) \;\;\; ; \;\;\; \frac{dt}{d\tau} = \frac{e}{1 - 2 M/r} \; ,
\eeq
where $M$ is the black hole mass, $e$ the energy per unit rest mass of the blob, $r$ the black hole-blob distance, $t$ the coordinate time, and $\tau$ the blob's proper time. These two equations can be solved for $dt/dr$ and integrated to obtain the coordinate time as a function of the orbital radius of the blob's COM,
\beq
t_{Sch}(r) = \int_{R_{ISCO}}^{r} \frac{e}{\left(1-\frac{2 M}{r'}\right)\sqrt{\frac{2M}{r'}-(1-e^2) }} dr' \; .
\eeq
If the blob is ejected out of a Kerr black hole, a similar set of equations can be solved to obtain its COM motion in the equatorial plane,
\beq
t_{Kerr}(r) = \int_{R_{ISCO}}^{r} \frac{e}{\Delta}   \frac{r'^2+a^2+\frac{2a^2M}{r'} }{\sqrt{e^2 + \frac{2 M a^2 e^2}{r'^3} + \frac{a^2 e^2}{r'^2} - \frac{\Delta}{r'^2} }} dr' \; ,
\eeq
where $a$ is the black hole's spin parameter and $\Delta(r) \equiv r^2 - 2 M r + a^2$. In general, there is no reason for the blob to be ejected in the equatorial plane of the black hole, and in fact blobs should preferentially be ejected along the spin axis. But, as shown in Figure \ref{spin}, the effect of the black hole spin is weak. At $t=10 M$, the trajectory of a blob with $e=2$ launched from an $a=0.999$ black hole is only $0.36 M$ apart from one launched from an $a=0$ black hole.

\section{Simulated observations}

In simulating what would be seen by radio interferometers, we project the COM motion of the blob to the sky plane far from the black hole. We utilize the \emph{geokerr} code \citep{Dexter} to trace rays from the observer plane located at infinity to the position of the blob. The coordinates $(x, y)$ parameterize positions in this observer plane. The Fourier transform of this plane yields the visibility of a radio interferometer.

\begin{figure}
\centering
\includegraphics[width=0.4\textwidth]{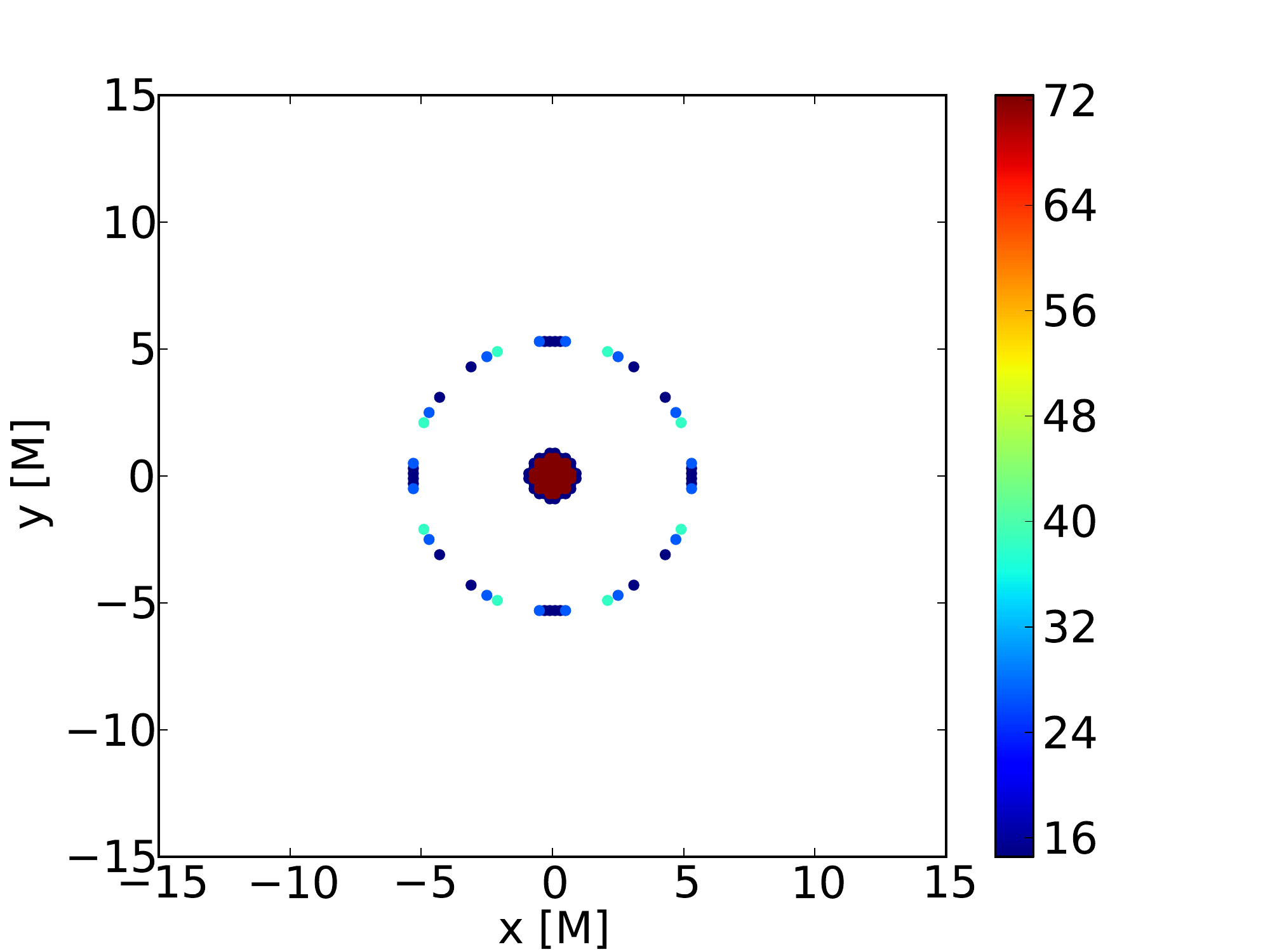}
\includegraphics[width=0.4\textwidth]{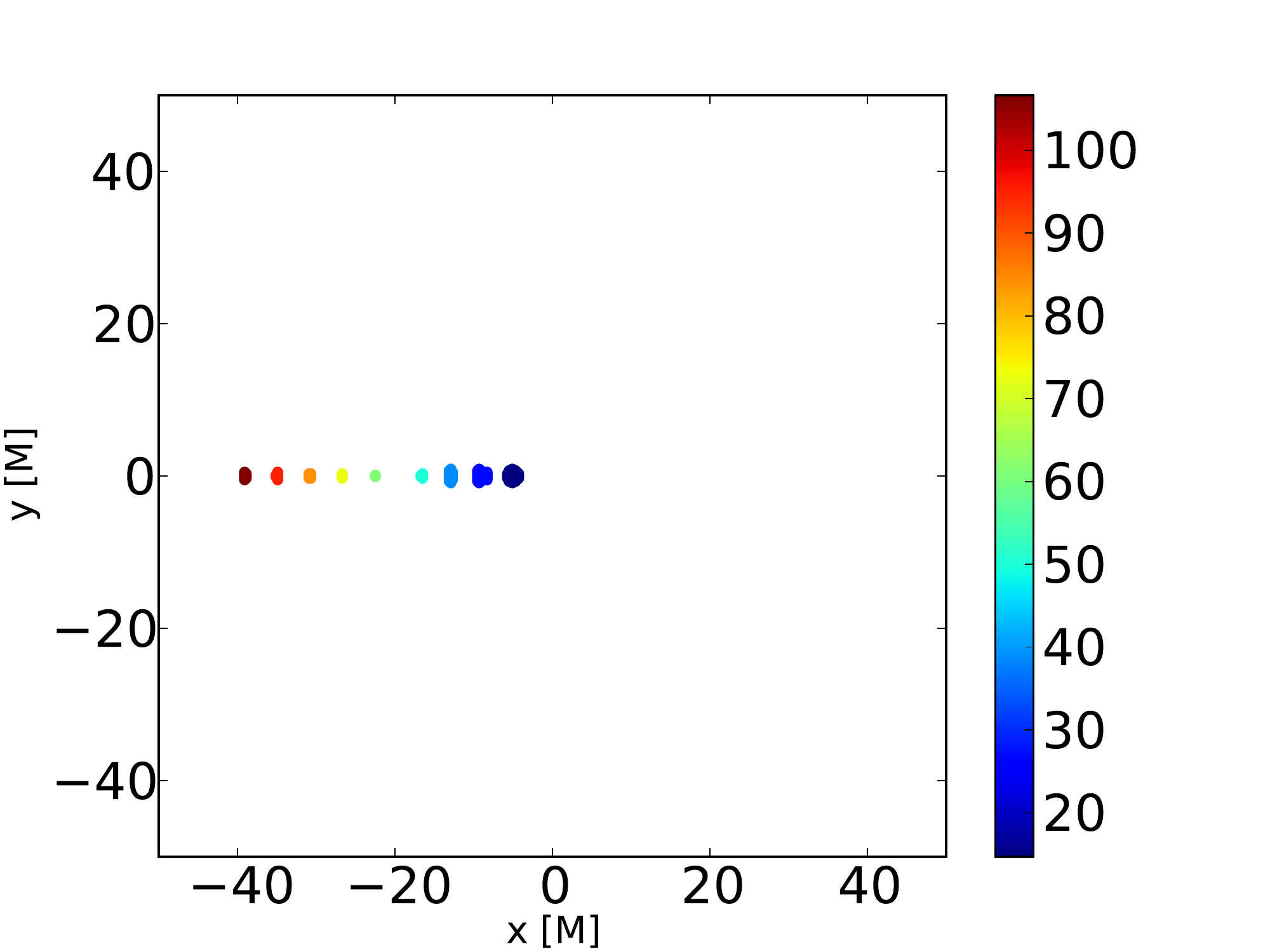}
\caption{Blobs with $e=10$ and radius $M$ launched with $\theta = 0$ and  $\theta = \pi / 8$ as seen in the observer plane with the black hole located at (0,0). The observer's time axis (in units of $M$) is indicated by the color bar. For a blob moving with $\theta=0$, the image is briefly lensed into a ring. The eccentricity of this ring can be used to test the no-hair theorem.}
\label{blobs}
\end{figure}

The blob itself is modeled as a small sphere that is emitting isotropically in its rest frame.  The result for blobs with velocity vectors at angles $\theta=0$ and $\theta=\pi /8 $ away from the observer are presented in Figure \ref{blobs}. For a blob moving along the $\theta=0$ axis, the image is briefly lensed into a ring with radius $R_{ring}\sim5 M$. Previous calculations by \cite{photonring} showed that the eccentricity of this ring is not sensitive to the spin of the black hole (except for $a\approx 1$), but is very sensitive to the black hole's quadrupole moment. Thus, if detected, the ring can be used as a test of the no-hair theorem. As the ring only appears when the blob is still close to the black hole, its lifetime is short ($\sim 40 M$ for a blob with $e=10$, but longer for slower moving blobs). It is therefore necessary to have temporal resolutions on minutes timescale to detect the ring. 

In addition, if the motion is fast enough and is launched at a small angle relative to the observer, the apparent trajectory can appear superluminal \citep[e.g.][]{Rees}. Close to the black hole, this apparent superluminal motion will be obscured by the bright photon ring. Thus, the detection of superluminal motion will require either waiting for the ring to dim or a manual removal of the ring. 

\begin{figure}
\centering
\includegraphics[width=0.4\textwidth]{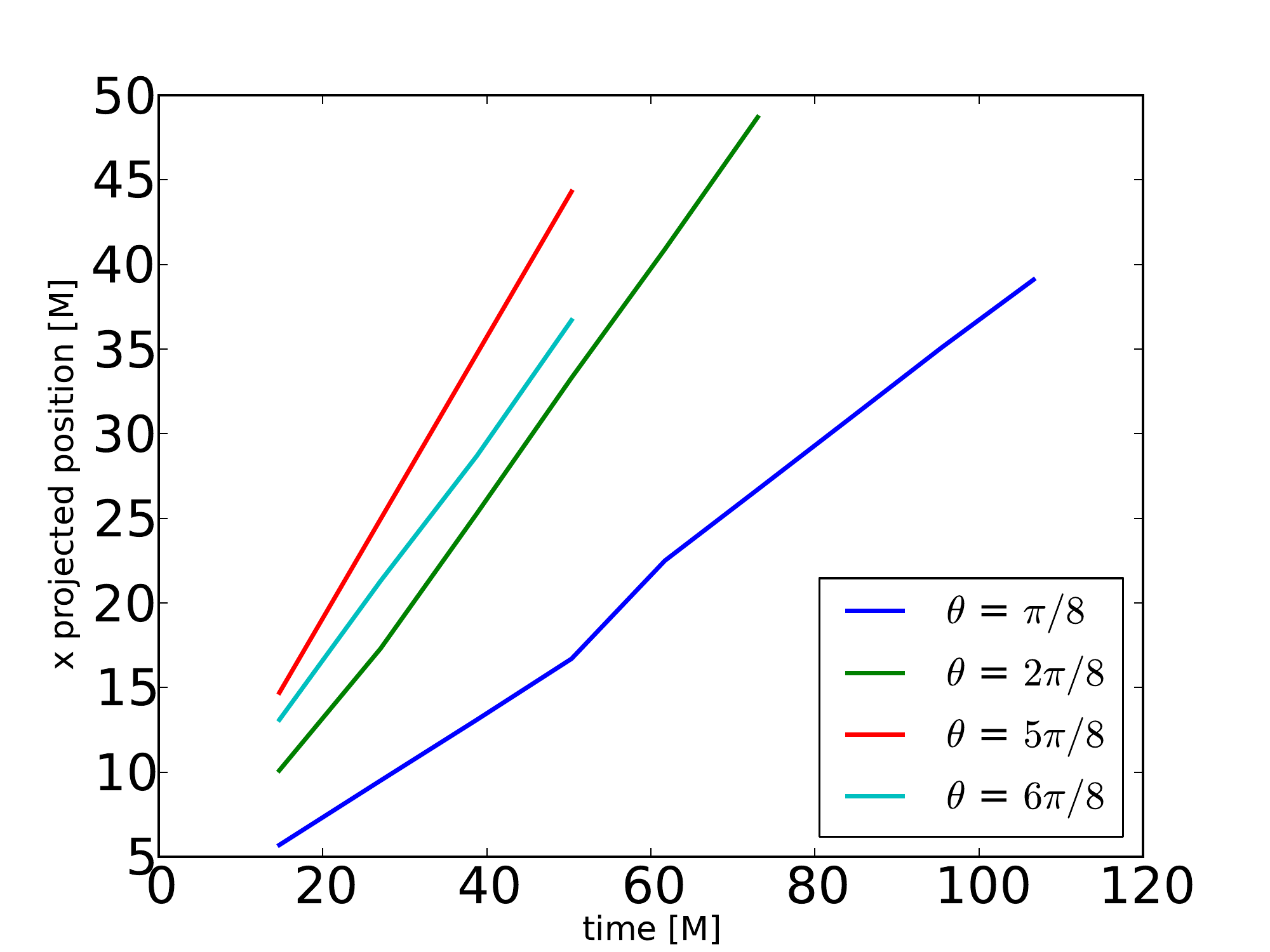}
\caption{The projected position of blobs with $e=10$ launched at a variety of angles versus observer time. }
\label{xproj}
\end{figure}

The projected distance as a function of observed times, shown in Figure \ref{xproj}, can be compared with observations to determine the presence of non-gravitational forces (e.g. due to magnetic fields or hydrodynamic friction on background gas). In addition, it can be used to constrain gravitational theories that predict changes on the orbit of test particles close to a black hole \citep[e.g.][]{giddings}.

\section{Tidal effects}
If the forces holding the blob together are much smaller than the tidal gravitational forces, the blob will be tidally sheared. The magnitude of this tidal shear depends on the black hole's mass and spin, and thus can be used to probe the black hole metric. Under the approximation that the force per unit mass keeping the blob together is $\ll (2 M R/r^3) $, where $R$ is the radius of the blob, the elements of the blob can be treated as if they are moving along geodesics. 

If the blob is small, we can define the geodesic deviation vector $\xi^\alpha$ between the geodesic followed by the particle at the center of the blob and the different geodesic followed by particles at the blob's edge by,
\beq
\xi^\alpha = \frac{\partial x^\alpha}{\partial s} \; ,
\eeq
where $s$ is the parameter indexing neighboring geodesics. We can calculate the rate of change of $\xi^\alpha$ with respect to the affine parameter of the geodesic,
\begin{align}
\frac{d}{d \tau} \xi^{\alpha} &= u^{\beta}\nabla_\beta \xi^\alpha - \Gamma^\alpha_{\beta \gamma} \xi^{\gamma} u^\beta 
\\&= \xi^\beta \nabla_\beta u^\alpha - \Gamma^\alpha_{\beta \gamma} \xi^{\gamma} u^\beta \; ,
 \end{align}
where we have used the identity \citep{Poisson},
\beq
u^\beta \nabla_\beta \xi^\alpha = \xi^\beta \nabla_\beta u^\alpha \; 
\eeq
which is valid for geodesic deviation vectors. Writing explicitly,
\begin{align}
\xi^{\beta}\nabla_\beta u^\alpha = \xi^\beta \frac{\partial u^\alpha}{\partial x^\beta} + \Gamma^\alpha_{\beta \gamma } u^\gamma \xi^\beta \; ,
\end{align}
yields 
\begin{align} \label{eq:dev}
\frac{d}{d \tau} \xi^\alpha = \xi^\beta \frac{\partial u^\alpha}{\partial x^\beta} \; .
\end{align}
The four velocity of a blob ejected from a Schwarzschild black hole with negligible angular momentum is:
\beq
u^\alpha = \left( \frac{e}{1 - \frac{2 M}{r}}, -\sqrt{\frac{2M}{r}-(1-e^2)},0,0 \right)^\alpha \; ,
\eeq
For relative motion between particles at the center of the blob and particles at the edge of the blob in the radial direction:
\beq \label{eq:xi}
\xi^\alpha = (0, R, 0, 0)^\alpha \; .
\eeq
Plugging equation (\ref{eq:xi}) into equation (\ref{eq:dev}) gives:
\beq \label{eq:dR/dlambda}
\frac{1}{R}\frac{d R}{d\lambda} = - \frac{M}{r^2 \sqrt{-1 + e^2 + \frac{2 M}{r}}} \; .
\eeq
Note that substituting $t$ for $\lambda$ in equation (\ref{eq:dR/dlambda}), then taking a derivative with respect to $t$ with $M/r \to \infty$ reproduces the tidal acceleration of Newtonian gravity: $a_{tidal} \sim M R/r^3$. 

Substituting the orbital radius $r$ in place of $\lambda$ in equation (\ref{eq:dR/dlambda}) and integrating, we get:
\begin{align}
\int_{R_0}^{R} \frac{d R'}{R'} =   -\int_{r_0}^{r} \frac{M dr'}{r'^2 \left(-1 + e^2 + \frac{2 M}{r'}\right) }  \;,
\end{align}
where $R_0 \ll r$ is the initial size of the blob and $r_0$ the starting orbital radius of the blob. Assuming that the blob is ejected from the ISCO radius, $r_0 = 6 M$ for $a=0$, we obtain:
\beq \label{eq:R/Ro}
\frac{R}{R_0} =  \left[ \frac{(-2 + 3 e^2)r}{6 M +3 (e^2 -1) r} \right]^2 \; .
\eeq
This change in radius is in principle observable, and can therefore be used to find the mass of the black hole if $e$ is inferred from the COM trajectory. The constant $e$ can be inferred far away from the black hole where it obeys $e = 1/\sqrt{1-v_{COM}^2}$, where $v_{COM}$ is the COM velocity of the blob at $r\gg M$. Figure \ref{tidal} shows the radial growth factor for blobs with specific energy $e=1.0001,\;1.001, \;1.01, \;\text{and}\;10$. Because blobs of smaller $e$ spend more time close to the black hole, the tidal effect is larger the closer $e$ is to unity. In the case of $e\sim1$, one can get a growth factor of $R/R_0 \sim 10$ at $r=1000M$. This is a change that is observable by the EHT. In general, one can also compute the relative motion between the center and the edge of the blob in the $\hat{\phi}$ and $\hat{\theta}$ direction via an analogous calculation.

\begin{figure} \label{tidal}
\centering
\includegraphics[width=0.4\textwidth]{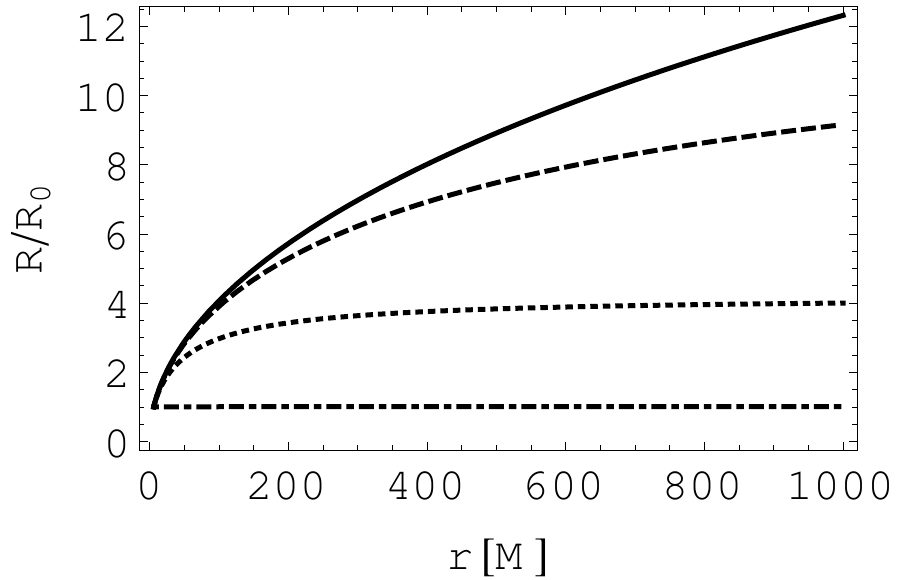}
\caption{The growth factor of the blob radius due to gravitational tide as a function of distance from the black hole for a blob moving with negligible angular momentum. The blob's specific energy is $e=1.0001,\;1.001,\; 1.01, \; \text{and} \;10$ for the solid, dashed, dotted lines, and dot-dashed lines, respectively.}
\label{tidal}
\end{figure}

\begin{figure}
\centering
\includegraphics[width=0.4\textwidth]{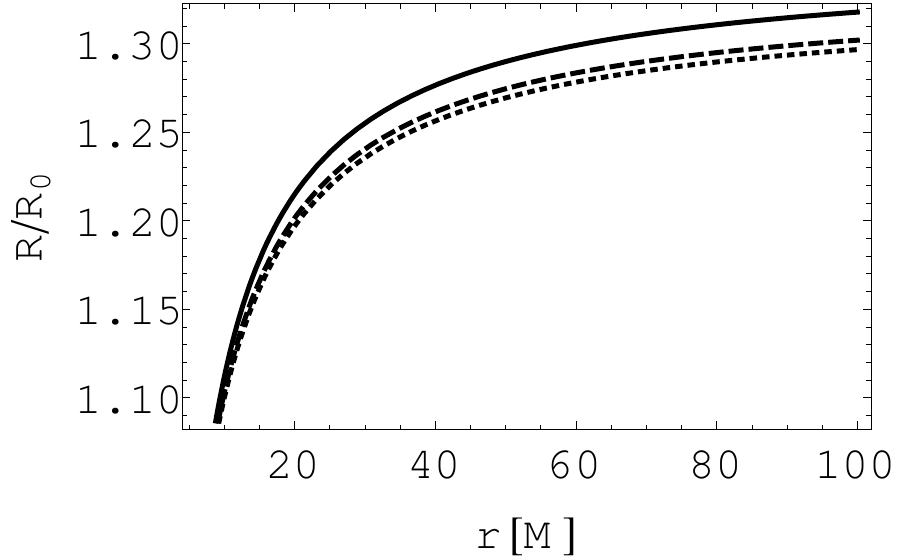}
\caption{The growth factor of the blob radius as a function of distance from a spinning black hole for a blob trajectory with a negligible angular momentum. The black hole's spin is $a=0,\;0.5,\;\text{and}\;1$ for the solid, dashed, and dotted lines, respectively. The blob energy is $e=1.2$ for all curves.}
\label{tidal_spin}
\end{figure}

We can extend this calculation to the case of a spinning black hole with a blob moving radially in the equatorial plane. For this configuration, the relevant components of $u^\alpha$ are,
\begin{align}
u^t &= \frac{e}{\Delta} \left( r^2 + a^2 + \frac{2 a^2 M}{r}  \right) \; ,
\\ u^r &= \sqrt{e^2 + \frac{2 M}{r^3} (a e)^2 + \frac{a^2 e^2}{r^2} - \frac{\Delta}{r^2}  } \; .
\end{align}
Again we adopt, 
\beq
\xi^\alpha = (0, R, 0, 0)^\alpha \; .
\eeq
Performing an analogous calculation as in the $a=0$ case, we obtain,
\begin{align} 
\begin{split}
&\frac{R}{R_0}= 
\\& \frac{R_0 \sqrt{a^2 \left(-3+4 e^2\right) M+36 \left(-2+3 e^2\right) M^3} r^{3/2}}{6 M^{3/2} \sqrt{3 r^2 \left[2 M+\left(-1+e^2\right) r\right]+3 a^2 \left[-r+e^2 (2 M+r)\right]}}  \; .
\end{split}
\end{align}
If the mass of the black hole and the blob energy $e$ are known, this equation can be used to measure the spin of the black hole. Figure \ref{tidal_spin} shows the growth factor $R/R_0$ for blobs with dimensionless spin parameter $a=0,\;0.5,\;\text{and} \;1$. The effect of spin is weak, and its measurement would be challenging.

\section{Conclusion}
We have shown that observations of ejected plasma blobs from the supermassive black holes Sgr A* and M87, can be used to constrain the spacetime near these black holes. There are two pieces of information that can be obtained from these observations: the blob's trajectory and the tidal effects on the blob's shape. 

The trajectory of the blob can be used to limit the presence of non-gravitational forces around the black hole or to constrain theories of gravity that predict anomalies in the orbit of test particles in the vicinity of black holes \citep[e.g.][]{giddings}. If a photon ring is detected, its eccentricity could be used as a test of the no-hair theorem. Furthermore, observations of the tidal stretching of the ejected blob can be used to determine both the mass and spin parameter of the black hole.

\section{Acknowledgment}
This work was supported in part by NSF grant AST-1312034.


\begin{thebibliography}{99}
\bibitem[\protect\citeauthoryear{Chandrasekhar}{1983}]{Chandra} Chandrasekhar, S. 1983, The Mathematical Theory of Black Hole,
Oxford University Press, Oxford
\bibitem[\protect\citeauthoryear{Cordes et al.}{2002}]{c10} Cordes, J. M. et al. 2002,
New AR, 48, 1413
\bibitem[\protect\citeauthoryear{Dexter and Agol}{2009}]{Dexter} Dexter J. and Agol, E. 2009,
apj, 696, 1616
\bibitem[\protect\citeauthoryear{Eisenhauer et al.}{2003}]{Dsgr} Eisenhauer F. et al. 2003,
ApJ, 597, L121
\bibitem[\protect\citeauthoryear{Gebhardt et al.}{2011}]{DM87} Gebhardt K. et al. 2011,
ApJ, 729, 119G
\bibitem[\protect\citeauthoryear{Genzel et al.}{2003}]{Genzel} Genzel, R. et al. 2003,
Nature, 425, 934
\bibitem[\protect\citeauthoryear{Ghez et al.}{2008}]{ghez} Ghez, A. M. et al. 2008,
ApJ, 689, 1044
\bibitem[\protect\citeauthoryear{Giddings}{2014}]{giddings} Giddings, S. B.  2014,
PhysRevD, 90, 12
\bibitem[\protect\citeauthoryear{Gillessen et al.}{2009}]{massbh} Gillessen, S. et al. 2009,
ApJ, 692, 1075
\bibitem[\protect\citeauthoryear{Johannsen}{2012}]{a3} Johannsen T. 2012,
PASP, 124, 1133
\bibitem[\protect\citeauthoryear{Johannsen and Psaltis}{2010}]{photonring} Johannsen T. and Psaltis D. 2010,
ApJ, 718, 446J
\bibitem[\protect\citeauthoryear{Johnson et al.}{2014}]{Johnson} Johnson M. D. et al. 2014,
ApJ 
\bibitem[\protect\citeauthoryear{Kramer et al.}{2004}]{c11} Kramer, M. et al. 2004,
New AR, 48, 993
\bibitem[\protect\citeauthoryear{Liu et al.}{2012}]{c6} Liu K. et al. 2012,
ApJ, 747, 1
\bibitem[\protect\citeauthoryear{Lu et al.}{2014}]{a1} Lu R. S. et al. 2014,
ApJ, 788, 120
\bibitem[\protect\citeauthoryear{Pfahl and Loeb}{2004}]{c4} Pfahl E., Loeb A. 2004,
ApJ, 615, 253
\bibitem[\protect\citeauthoryear{Poisson}{2004}]{Poisson} Poisson, E. 2004, A Relativists's Toolkit,
Cambridge University Press, Cambridge
\bibitem[\protect\citeauthoryear{Mirabel et al.}{1992}]{mirabel} Mirabel, I. F. et al. 1992,
Nature, 358, 215M
\bibitem[\protect\citeauthoryear{Mirabel}{2002}]{mirabel2} Mirabel, I. F.  2002,
IAU Symposium, 214, 201
\bibitem[\protect\citeauthoryear{Mirabel}{2004}]{mirabel3} Mirabel, I. F.  2004,
ESA Special Publication, 552, 175
\bibitem[\protect\citeauthoryear{Psaltis et al.}{2014}]{a2} Psaltis D. et al. 2014,
arxiv:1411.454
\bibitem[\protect\citeauthoryear{Psaltis and Johannsen}{2011}]{c5} Psaltis D., Johannsen T. 2011,
JPhCS, 283, 2030P
\bibitem[\protect\citeauthoryear{Rees}{1966}]{Rees} Rees M. J. 1966,
Nature, 211, 468R
\bibitem[\protect\citeauthoryear{Walsh et al.}{2013}]{M87} Walsh, J. L. et al. 2013,
ApJ, 770, 86W
\end{thebibliography}
\end{document}